# Generation of J0 Bessel Beams with controlled spatial coherence features

Adrian Carbajal-Dominguez[1]*, Jorge Bernal[1], Alberto Martin-Ruiz[1] and Gabriel Martínez Niconoff[2]

[1]*Universidad Juarez Autonoma de Tabasco, Division Academica de Ciencias Basicas, Cunduacan, Tabasco, C.P. 86690, Mexico*

[2]*Instituto Nacional de Astrofísica, Óptica y Electrónica, INAOE, Grupo de Óptica Estadística, Apdo. postal 51 y 216, C.P. 72000, Puebla, Pue., México*

**adrian.carbajal@dacb.ujat.mx*

**Abstract:** An alternative method to generate J0 Bessel beams with controlled spatial partial coherence properties is introduced. Far field diffraction from a discrete number of source points on an annular region is calculated. The average for different diffracted fields produced at several rotation angles is numerically calculated and experimentally detected. Theoretical and experimental results show that for this particular case, J0 Bessel beam is a limit when the number of points tends towards infinity and the associated complex degree of coherence is also a function of the number of points.

## References and links

1. J. Durnin, "Exact solutions for nondiffracting beams. I. The scalar theory," J. Opt. Soc. Am. A **4**, 651-654 (1987).
2. Z. Bouchal, J. Wagner, and M. Chlup, "Self-reconstruction of a distorted nondiffracting beam," Optics Communications **151**, 207-211 (1998).
3. S.H. Tao and X. Yuan, "Self-reconstruction property of fractional Bessel beams," J. Opt. Soc. Am. A **21**, 1192-1197 (2004).
4. Y. Liu, C. Gao, M. Gao, and F. Li, "Coherent-mode representation and orbital angular momentum spectrum of partially coherent beam," Optics Communications **281**, 1968-1975 (2008).
5. G.J. Williams, H.M. Quiney, A.G. Peele, and K.A. Nugent, "Coherent diffractive imaging and partial coherence," Phys. Rev. B **75**, 104102 (2007).
6. F. Dubois, N. Callens, C. Yourassowsky, M. Hoyos, P. Kurowski, and O. Monnom, "Digital holographic microscopy with reduced spatial coherence for three-dimensional particle flow analysis," Appl. Opt. **45**, 864-871 (2006).
7. Z. Bouchal and J. Perina, "Non-diffracting beams with controlled spatial coherence," JOURNAL OF MODERN OPTICS **49**, 1673-1689 (2002).
8. H.T. Eyyuboglu, "Propagation and coherence properties of higher order partially coherent dark hollow beams in turbulence," Optics & Laser Technology **40**, 156-166 (2008).
9. K. Saastamoinen, J. Turunen, P. Vahimaa, and A. Friberg, "Spectrally partially coherent propagation-invariant fields," PHYSICAL REVIEW A **80**, (2009).
10. G.M. Niconoff, J. Ramírez San Juan, J.M. López, and P.M. Vara, "Incoherent convergence of diffraction free fields," Optics Communications **275**, 10-13 (2007).
11. R. Bracewell, *The Fourier Transform & Its Applications*, 3º ed. (McGraw-Hill Science/Engineering/Math, 1999).
12. J.W. Goodman, *Introduction to Fourier Optics*, 3º ed. (Roberts & Company Publishers, 2004).
13. M. Abramowitz and I.A. Stegun, *Handbook of Mathematical Functions: With Formulas, Graphs, and Mathematical Tables* (Dover Publications, 1965).
14. L. Mandel and E. Wolf, *Optical Coherence and Quantum Optics*, 1º ed. (Cambridge University Press, 1995).
15. W.H. Press, B.P. Flannery, S.A. Teukolsky, and W.T. Vetterling, *Numerical Recipes in C: The Art of Scientific Computing*, 2º ed. (Cambridge University Press, 1992).

## 1. Introduction

One of the first known methods to generate J0 Bessel beams consists of illuminating an annular slit [1]. Bessel beams are also known to be non-diffractive or diffraction free optical beams. This means that they do maintain a constant profile along propagation coordinate. They also have the surprising property of self-regeneration [2,3]. Recently, it has been a growing interest in the synthesis and characterization of Bessel and other optical beams with controlled coherence properties. The idea is to combine the benefits of partial coherence beams with non-diffracting ones in order to obtain better performances in applications such as free space telecommunications [4], imaging [5], microscopy [6], among others.

Others authors have reported the study of propagation of partially coherent beams in different conditions as in the atmosphere, considering turbulence or polychromatic waves [7-9]

The purpose of this work is to report how a J0 Bessel beam with controlled spatial partial coherence features can be generated. We use the far field form a mask composed of a finite number of light point sources equally spaced placed on an annulus of radius $R$. Then, the average of several of such fields produced at different angular positions is taken. Numerical and experimental results show that these optical fields have as limit the J0 Bessel beam as the number of points $n$ tends towards infinity. In order to corroborate the later, complex degree of coherence is studied as a function of the number of point sources. We believe that these results confirm the convergence of incoherent optical beams [10]

## 2. Theory

As a start, we consider the diffracted intensity produced by $n$ source points equally spaced placed on a circle of radius $R$. In this case, the amplitude transmittance is function of polar coordinates $r, \theta$ and an angular random variable $\Delta$ is of the form

$$t(r,\theta;\Delta) = \frac{\delta(r-R)}{\pi r} \sum_{p=1}^{n} \delta\left[\theta - \left(\frac{2\pi p}{n} - \Delta\right)\right], \qquad (1)$$

whose geometrical parameters are better explained in fig(1) and Dirac delta function has been written in polar coordinates [11], $n$ is the number of equally spaced point sources on the circle of radius $R$, $\Delta$ is a random rotating angle applied simultaneously to all point sources and it is treated as a random variable.

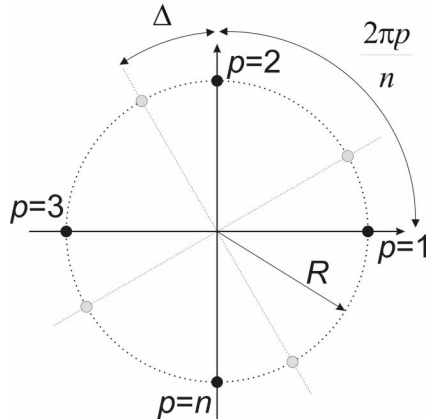

figure 1 Geometrical parameters. a) original angular position depending on the number of points . b) each point is rotated to angle $\Delta$ , but all points remain on a circular trajectory because the same radius $R$ is maintained constant.

The diffracted field is calculated in the context of angular spectrum of plane waves theory. In polar cylindrical coordinates, the angular spectrum of plane waves [12], has the form,

$$A(\rho,\varphi) = \frac{1}{\pi} \int_0^{2\pi} \int_0^\infty \delta(r-R) \sum_{p=1}^n \delta\left[\theta - \left(\frac{2\pi p}{n} - \Delta\right)\right] \exp[i2\pi r\rho \cos(\theta - \varphi)] dr d\theta \quad (2)$$

Performing the integral (2) leads to an angular spectrum depending only on radial coordinate $\rho$ on momentum space, as follows.

$$A(\rho,\varphi) = \frac{1}{\pi} \sum_{p=1}^n \sum_{q=-\infty}^{\infty} i^q J_q(2\pi R\rho) \exp\left[iq\left(\frac{2\pi p}{n} - \Delta - \varphi\right)\right], \quad (3)$$

where the Jacobi-Anger identity has been used [13]. Hence, the optical near field is of the form,

$$\phi(r,z;\Delta) = \int_0^{2\pi} \int_0^\infty A(\rho,\varphi) J_0(2\pi r\rho) \exp[i2\pi z\sqrt{1/\lambda^2 - \rho^2}] \rho d\rho d\varphi, \quad (4)$$

in polar cylindrical coordinates. Whereas for the far field, $\phi(r,z_\infty;\Delta) \approx A(r/z;\Delta)$ which is Fraunhofer diffraction for the angular spectrum context [14]. In this case, $\lambda$ is the wavelength, $z$ is the propagation coordinate.

The associated irradiance for one of these far fields is known to be $I(r,z_\infty;\Delta) = |\phi(r,z;\Delta)|^2$. Eq. (3) is valid for all $\varphi$, in particular we consider $\varphi = 0$ without any loss of generality and for the sake of clarity. Averaging all these irradiance distributions on the entire $\Delta$ ensemble, lead us to

$$\overline{I}(\Delta) = \langle I(r,z_\infty) \rangle_\Delta = \frac{1}{\pi^2} \left\{ J_0(2\pi Rr)^2 + 2\sum_{k=1}^\infty J_k(2\pi Rr)^2 \sum_{p=1}^n \sum_{q=1}^n \cos\left[\frac{2\pi k}{n}(p-q)\right] \right\}, \quad (5)$$

and the associated radial correlation function for such averaged intensity distributions with the $J_0$ Bessel function can be expressed as

$$\Gamma[\overline{I}(r;v), J_0(r)] = \int_0^\infty \overline{I}(r;n) J_0(r) dr \cdot \quad (6)$$

From here, it is possible to write the complex degree of coherence $\gamma$ in the form [14]

$$\gamma = \frac{\int_0^\infty \overline{I}(r;n) J_0(r) dr}{\left(\int_0^\infty \overline{I}(r;n)^2 dr\right)^{1/2} \left(\int_0^\infty J_0(r)^2 dr\right)^{1/2}} \quad (7)$$

*2.1 Numerical simulation*

In order to grasp a deeper understanding on the behavior of eq.(5), a numerical evaluation is performed. The Bessel functions are approximated by using polynomial approximation for the first 100 Bessel functions [13,15]. Cases ranging from $n = 2$ to 50 points are considered. In fig. (2) results are shown for some of these cases. It can be seen that as *n* is increased, eq.(5) resembles Bessel function of order zero. It was observed that the second term in eq.(5)

forms around $r = 0$ a zero value plateau which radius increases with $n$. Over this interval, the $J_0(2\pi Rr)^2$ term is completely dominant. Outside this interval, both terms interact and the result is a function that is dominated by the second term which is different from Bessel function of order zero.

In order to understand how well is the resemblance of these functions with $J_0(2\pi Rr)$, the complex degree of coherence $\gamma$ given in eq.(7) is calculated for several cases. Results are shown in fig. (3). In this case, $\gamma$ clearly increases towards 1 as $n$ tends towards infinity. The asymptotic value obtained for $\gamma$ is 0.994 with a standard error of $5 \times 10^{-3}$.

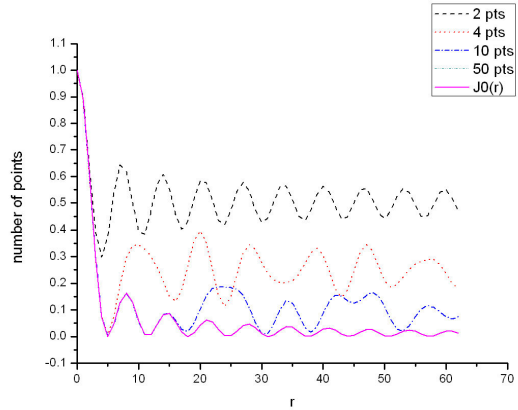

figure 2 Plot of eq. (5) for 2, 4, 10 and 50 points plus J0 function. As the number of points increases, the function tends towards the Bessel function, as can be seen for the case of 50 points.

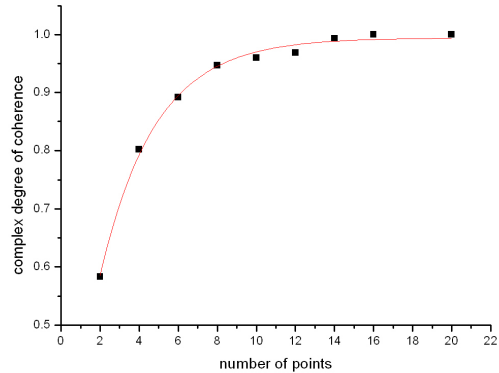

figure 3 Plot of $\gamma$ for different number of points. For $n \geq 20$ the complex degree of coherence is very near to 1. The asymptotic value obtained is 0.994 with a standard error of $5 \times 10^{-3}$.

A different but complementary approach is to numerically calculate far field diffraction for transmittance functions given in eq. (1). The program consists of three parts. First, the transmittance function with $n$ evenly distributed points is calculated. Next, a routine to rotate these *n* points to a certain angle is added. Then, a far field diffraction pattern is computed for each rotated transmittance by means of the standard 2D Fast Fourier Transform (FFT) algorithm [15], and a squared modulus is calculated in order to obtain the corresponding irradiance. Fifty different fields are generated at different $\Delta$ random rotation angles. Finally, the averaged intensity of all these fields is evaluated. Numerical results are shown in figure (2) for three particular cases out of 80 that have been studied in this way. The transversal intensity of the diffracted fields are plotted in fig. (4). It can be seen the same behavior as the observed for the plot presented in fig.(2) if one, for instance, observes the number of lobes for the same number of points. These numerical results coincide with each other, despite different assumptions were made for their calculation. Moreover,

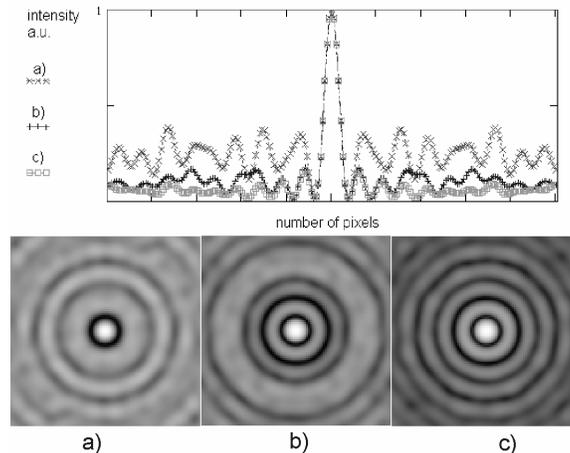

figure 4. Numerical results for different cases. a) 4 points case, b) 10 points case, c) 20 points case. Associated transversal normalized intensity distribution is plotted above. As the number of coherent points increases, the image contrast increases and the intensity distribution tends towards a squared J0 Bessel function. The behavior of the lobes coincide with fig.(2).

## 4. Experiment

Transmittance functions with different number of points are computer generated and printed on overhead projector transparencies. Each transparency is mounted on a PC controlled rotating stage. An HeNe laser $\lambda = 632.8 nm$ is used to create a coherent plane wave to illuminate them. Then, a positive doublet $f = 85.0 cm$ is used as a transformer lens. A black and white CCD camera with no lens is placed at Fourier´s plane. When the transmittance function is set to rotate, the video capture is made randomly and 900 frames are recorded for each case. Then, the associated average is calculated over the entire number of captured images. Results are shown in fig. (5) for three different cases.

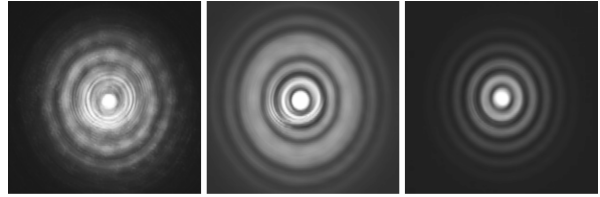

figure 5 Experimental results for different cases a) 4 points b) 10 points c) 50 points. The intensity distributions are in good agreement with results obtained numerically in fig. (4), the number of minima and maxima of intensity are the same. The corresponding $\gamma$ values are: 0.444, 0.649 and 1.

Averaged images have been contrast corrected in order to obtain a better visual comparison with previously obtained numerical results. It can be observed that the experimental results are in very good agreement with the numerical results, as for the coincidence in the number of lobes for each case and for the axial intensity distribution. In both cases, as was expected, the resulting field tended towards a squared J0 Bessel function when the number of points was increased, as occurred in the case of an annular slit transmittance. Hence, the maximum intensity of a J0 Bessel beam is the result of the coherent contribution of an infinite number of points on an annular slit. As the number of such points decreases, the intensity is spatially redistributed generating an optical field with a weakened contrast and a less focused energy distribution. The complex degree of coherence given in eq. (7) is also computed for differents averaged fields, and results for 4, 10 and 50 point are 0.444, 0.649 and 1, respectively. This also confirms the tendency depicted in fig. (3) where the complex degree of coherence tends towards 1 as the number of points increases to infinity.

## 5. Conclusions

In conclusion, a method that allows the generation of J0 Bessel like beams with controlled spatial partial coherence properties has been introduced. The method is based on the average of far-field intensity distributions produced by *n* equally spaced point sources placed on a circle of radius *R* over an ensemble of optical fields that have been rotated to different random angles. Numerical and experimental results show that this incoherent average tends towards the intensity distribution of a J0 Bessel beam as the number of points is increased. Meanwhile, for the intermediate cases, the resulting averaged optical fields are characterized by a complex degree of coherence which is directly related to he number of points. We believe that these fields can be thought as partially coherent J0 Bessel beams.